\def\be{\begin{equation}}
\def\ee{\end{equation}}
\def\ba{\begin{array}}
\def\ea{\end{array}}

\documentclass[preprint,showpacs,amsmath]{revtex4}
\usepackage{amsmath}
\usepackage{float}
\usepackage[dvipsnames, svgnames, x11names]{xcolor}
\usepackage{graphicx} 
\usepackage{epstopdf}
\usepackage{lipsum}
\usepackage{caption}
\usepackage{pifont}
\usepackage{amssymb}
\usepackage{algorithm }
\usepackage{algorithmic}
\usepackage{amsfonts}

\def\qed{\leavevmode\unskip\penalty9999 \hbox{}\nobreak\hfill
     \quad\hbox{\leavevmode  \hbox to.77778em{%
               \hfil\vrule   \vbox to.675em%
               {\hrule width.6em\vfil\hrule}\vrule\hfil}}
     \par\vskip3pt}

\newtheorem{theorem}{Theorem}

\newtheorem{lemma}{Lemma}

\allowdisplaybreaks[4]

\input amssym.def

\begin{document}

\title{Criteria for SLOCC and LU Equivalence of Generic Multi-qudit States}
\author{Jingmei Chang$^{1}$, \ Naihuan Jing$^{2, \dag}$, Tinggui Zhang$^{3}$\\ 
$^{1}${Department of Mathematics, Shanghai University}\\  
$^{2}${Department of Mathematics, North Carolina State University}\\ 
$^{3}${School of Mathematics and Statistics, Hainan Normal University} \\
$^\dag$
Corresponding author. E-mail: jing@ncsu.edu}
\bigskip

\begin{abstract}
In this paper, we study the stochastic local operation and classical communication (SLOCC) and local unitary (LU) equivalence for multi-qudit states by mode-$n$ matricization of the coefficient tensors. We establish a new scheme of using the CANDECOMP/PARAFAC (CP) decomposition of tensors to find necessary and sufficient conditions between the mode-$n$ unfolding and SLOCC\&LU equivalence for pure multi-qudit states. For multipartite mixed states, we present a necessary and sufficient condition for LU equivalence and necessary condition for SLOCC equivalence.

\end{abstract}

 \maketitle

\bigskip

\section{Introduction}
Quantum entanglement is one of the essential features of quantum mechanics that plays an important role in quantum information processes \cite{cy,qp2} quantum teleportation \cite{qt1,qt2,qt3,qt4}, quantum cryptography \cite{qc1,qc2}, and quantum computation \cite{qic1,qic2,qic3}, dense coding \cite{dc} and so on.
Since entanglement is closely related to non-local properties of a state, it cannot be affected by local quantum operations realized by the group of
stochastic local operations and classical communication  (SLOCC) operations \cite{entang1,entang2}.
It is therefore imperative to classify different types of entanglement for a given quantum state.
In recent years, studies have been conducted on the connections between invariant theory and entanglement by qualitative and quantitative characterization. The SLOCC and local unitary (LU) equivalence of density matrices are helpful to classify entanglement at the local level.  It is well known that, within the same SLOCC class, the states are able to perform the same QIT-tasks \cite{entang2,qts2}, and under local unitary transformations, entanglement of two states remains completely equivalent.

It is thus important to find equivalence conditions under SLOCC and LU for any two states. There have been a lot of research on
the SLOCC entanglement classification \cite{entang2,instu2,qts2,instu4,instu5,instu6,instu7,instu8,instu9}. Classification of three-qubit systems into six types was done in \cite{entang2} for pure states and then generalized to three-qubit mixed states in \cite{instu2}. For four-qubit pure states, there are nine distinct SLOCC classes using group theory \cite{qts2}.
In 2012, $n$-qubit pure states were classified by the ranks of the coefficient matrices in \cite{instu4}. Wang et al's method \cite{instu5,instu6} generalized the conclusion to higher-dimensional multipartite pure states under SLOCC transformation. 
Ref. \cite{instu7} proposed a new approach to the geometry of the four-qubit entanglement classes depending on parameters. Turner \cite{instu8} gave new upper bounds on the degrees of the invariants for certain complete set of the n-qubit SLOCC invariants. The method of \cite{instu9} helped distinguish certain SLOCC classes within the null-cone. There were also many new results and approaches recently. For instance, Zangi proposed a practical classification scheme \cite{Zangi} for the four-partite entangled states by transforming a four-partite state into a triple-state set composed of two tripartite and one bipartite states to classify any multipartite entangled state. Shi \cite{Xian} considered the trivial stabilizer group for an $n$-qubit symmetric pure state. In 2022, Huang \cite{Huang} gave a mathematical framework for describing entanglement quantitatively and qualitatively by using rows or columns of the coefficient matrices for multipartite qudit states, and so on.

It is known that LU-equivalence classes are included in SLOCC-equivalence classes \cite{entang2}. The equivalence under LU can be done by the Schmidt decomposition for pure bipartite states \cite{cy}. The second-named  author presented a matrix method of
fixed point subgroup and tensor decomposability to characterize LU equivalence of
bi-partite states in joint work \cite{FeiJing}. Later, Kraus \cite{purenb} gave a general method to determine two LU equivalent $n$-qubit pure states. For multipartite pure states, Liu \cite{purend} generalized the method to higher dimensional case and Verstraete \cite{puredi} studied the classification under LU. For multipartite mixed states, a necessary and sufficient criterion of LU equivalence was obtained based on the matrix realignment \cite{mixmr}. A criterion for n-qubit mixed states based on Bloch representation was represented in \cite{nbbre}. There are also some results to determine local unitary equivalence by invariants, notably Makhlin \cite{poinva} gave a complete set of 18 polynomial invariants of two-qubit mixed states for the local unitary equivalence, Turner \cite{luinv} derived a complete set of invariants to distinguish any two distinct orbits, and \cite{Jing, Jing2, Sun} proposed a new set of invariants and
generalized some of these results. Nevertheless our knowledge on classification of multi-qudit states is still limited and further study is needed.

In this article, we consider necessary and sufficient conditions of SLOCC and LU equivalence for any {state} in an arbitrary-dimensional multipartite system by using the  mode-$n$ matricization of the coefficient tensors and its CP decomposition. The CP decomposition is an optimal approximation of
the general tensor in terms of special monomial tensors of lower ranks, which provides a means to study the relationship between the tensor and its n-mode unfolding. Our method offers a new approach to analyze SLOCC and LU equivalence. In Sec.\uppercase\expandafter{\romannumeral 2}, we briefly recall the CP decomposition of a tensor, the criterion for decomposition of block invertible matrix into tensor products of invertible matrices, and discuss some computation formulas. In Sec.\uppercase\expandafter{\romannumeral 3}, we mainly give some necessary and sufficient conditions for the SLOCC and LU equivalence of multipartite pure states and mixed states based on the mode-$n$ matricization of the coefficient tensors. We then use specific example to verify these criteria. The paper then ends with  the conclusions in Sec.\uppercase\expandafter{\romannumeral 4}.

\section{Preliminaries and Notations}

A tensor is a multidimensional array. In general, let $V_1, \cdots, V_N$ be vector spaces over $\mathbb F$ ($\mathbb F=\mathbb C$ or $\mathbb R$) with prescribed coordinate systems, then
a general vector in the tensor product $V_1\otimes\cdots\otimes V_N$ provides obvious configuration of order $N$ tensor.
Such a tensor $\mathcal{X}$ $(\in\mathbb F^{n_1\times\cdots\times n_N})$ with elements $x_{i_{1}\cdots i_{N}}(\in\mathbb{F})$ is said to be of an $N$-way or $N$th-order tensor.
For example, a first-order tensor is a vector, 
a second-order tensor is a matrix, 
and when the order is more than two we refer them as
higher-order tensors $\mathcal{X}$.
There are two ways to represent tensors. One is through fibers, which are higher-order analogues of matrix rows and columns, defined by fixing every index but one, i.e., $\mathbf{x}_{:jk}$, $\mathbf{x}_{i:k}$, and $\mathbf{x}_{ij:}$. The other is by slices, which are two-dimensional sections of a tensor, defined by fixing all but two indices, i.e., $X_{i::}$, $X_{:j:}$, and $X_{::k}$.

Matricization (unfolding or flattening) is the process of rearranging the elements of an $N$-way array into a matrix and is 
also called realignment in physics literature \cite{Jing,Tencp}.
Arranging the mode-$n$ fibers as the columns of the resulting matrix is called the mode-$n$ matricization (unfolding or flattening) of a tensor $\mathcal{X}\in\mathbb{R}^{\alpha_1\times \alpha_2\times\cdots\times \alpha_N}$, denoted by $X_{(n)}$ \cite{Tencp}.
The tensor element $x_{i_1,i_2,\dots,i_N}$ is mapped to the matrix entry $X_{(n)}(i_n,j)$ with
\begin{equation*}\label{rank1}
    j=1+\sum^{N}_{\substack{k=1 \\ k\neq n}}(i_{k}-1)\beta_k,
\end{equation*}
where $\beta_k=\prod\limits^{k-1}_{\substack{m=1 \\ m\neq n}}\alpha_{m}$.

For instance, if the frontal slices of a third-order tensor $\mathcal{X}\in\mathbb{R}^{3\times2\times2}$ are
$X_1=\left(\begin{smallmatrix}
     1 & 2\\
     3 & 4\\
     5 & 6
\end{smallmatrix}\right)$ and $X_2=\left(\begin{smallmatrix}
     7 & 8\\
     9 & 10\\
     11 & 12
\end{smallmatrix}\right)$, then the three mode-$n$ matricizations are
$X_{(1)}=\left(\begin{smallmatrix}
     1 & 2 & 7 & 8\\
     3 & 4 & 9 & 10\\
     5 & 6 & 11 & 12
\end{smallmatrix}\right)$, $X_{(2)}=\left(\begin{smallmatrix}
     1 & 3 & 5 & 7 & 9 & 11\\
     2 & 4 & 6 & 8 & 10 & 12
\end{smallmatrix}\right)$, and $X_{(3)}=\left(\begin{smallmatrix}
     1 & 3 & 5 & 2 & 4 & 6\\
     7 & 9 & 11 & 8 & 10 & 12
\end{smallmatrix}\right)$.

For a general tensor $\mathcal X$, if $\mathcal X$ can be written as a sum of component rank-one tensors, then such a decomposition is called a 
canonical polyadic decomposition (CP decomposition) for $\mathcal X$.
The rank $rank(\mathcal{X})$ of a tensor $\mathcal{X}$ is defined as the smallest number of rank-one tensors summing up to $\mathcal{X}$ \cite{1rank,2rank}. It is known that
any tensor $\mathcal{X}$ always admits a CP decomposition \cite{qi}.

Let $\mathcal{X}=(x_{i_1i_2\cdots i_N})\in\mathbb{R}^{k_1\times k_2\times\cdots\times k_{N}}$ be a 
 rank $R$ tensor, its CP decomposition can be written as
\begin{equation}\label{NX}
\mathcal{X}=\sum^{R}_{r=1}\mathbf{a}^{(1)}_{r}\circ\mathbf{a}^{(2)}_{r}\circ\cdots\circ\mathbf{a}^{(N)}_{r}=[\![{A_{1}},{A_{2}},\cdots,{A_{N}}]\!],
\end{equation}
where ``$\circ$'' represents the vector outer product, and $\mathbf{a}^{(i)}_{r}\in\mathbb{R}^{k_i}$ for $i=1,\dots,N$. Thus the elements of $\mathcal{X}$ in (\ref{NX}) are
\begin{equation}\label{3CE}
    x_{i_1i_2\cdots i_N}=\sum^{R}_{r=1}a^{(1)}_{i_1r}\cdots a^{(N)}_{i_Nr},
\end{equation}
and the factor matrices are composed of the rank-one components, i.e., $A_i=[\mathbf{a}^{(i)}_{1}, \ \mathbf{a}^{(i)}_{2}, \ \cdots \ \mathbf{a}^{(i)}_{R}]$. Clearly
the mode-$n$ matricization \cite{Tencp} of the $\mathcal{X}$ is given by
\begin{equation}\label{nmatricization}
    X_{(n)}=A_{n}(A_{N}\odot\cdots\odot A_{n+1}\odot A_{n-1}\odot \cdots\odot A_{1})^\mathrm{t},
\end{equation}
where $A_{n}\in\mathbb{R}^{k_n\times R}$, $n=1,2,\cdots,N$, and ``$\mathrm{t}$'' stands for the transpose. 
By computation, the CP decomposition obviously also exists in the complex field $\mathbb{C}$.

There is a natural action of $\mathrm{GL}_{n_1}\otimes\cdots\otimes\mathrm{GL}_{n_N}$ on $n_1\times\cdots\times n_N$ tensors as follows: for $\mathcal X=(x_{i_1\dots i_N})$,
\begin{equation}\label{tens}
\left((A_1\otimes \cdots \otimes A_N)\mathcal X\right)_{i_1i_2\dots i_N}=\sum\limits_{k_1,k_2,\dots, k_N} a_{i_1k_1}a_{i_2k_2}\cdots a_{i_Nk_N}x_{k_1k_2\dots k_N}
\end{equation}

We also need the following well-known products:

(i) The Kronecker product $A\otimes B$ is a matrix of size $(\alpha\gamma)\times(\beta\delta)$ defined by $A\otimes B=[\mathbf{a}_1\otimes\mathbf{b}_1, \ \mathbf{a}_1\otimes\mathbf{b}_2, \ \mathbf{a}_1\otimes\mathbf{b}_3, \ \cdots, \ \mathbf{a}_{\beta}\otimes\mathbf{b}_{\delta-1}, \ \mathbf{a}_{\beta}\otimes\mathbf{b}_{\delta}]$ for $A\in\mathbb{R}^{\alpha\times\beta}$, $B\in\mathbb{R}^{\gamma\times\delta}$, where $\mathbf{a}_{i}$ (or $\mathbf{b}_{i}$) is the column vector of matrix $A$ (or $B$).

(ii) The Khatri-Rao product \cite{khatri} $A\odot B$ is a matrix of size $(\alpha\beta)\times\gamma$ defined by
$A\odot B=[\mathbf{a}_1\otimes\mathbf{b}_1, \ \mathbf{a}_2\otimes\mathbf{b}_2, \cdots, \ \mathbf{a}_{\gamma}\otimes\mathbf{b}_{\gamma}]$ for $A\in\mathbb{R}^{\alpha\times\gamma}$, $B\in\mathbb{R}^{\beta\times\gamma}$, where $\mathbf{a}_{i}$ (or $\mathbf{b}_{i}$) is the column vector of matrix $A$ (or $B$).

If $\mathbf{a}$ and $\mathbf{b}$ are vectors, then the Khatri-Rao and Kronecker products are identical, i.e., $\mathbf{a}\otimes\mathbf{b}=\mathbf{a}\odot\mathbf{b}$.

(iii) The Hadamard product is a matrix of size $\alpha\times\beta$
defined by $A\ast B=(a_{ij}b_{ij})\in\mathbb R^{\alpha\times\beta}$ for $A,B\in\mathbb R^{\alpha\times\beta}$.

The following properties are well-known \cite{proper,khatri}:
$(A\otimes B)(C\otimes D)=AC\otimes BD$, $(A\otimes B)^{\dag}=A^{\dag}\otimes B^{\dag}$, $(A\odot B)^{t}(A\odot B)=A^{t}A\ast B^{t}B$, $(A\odot B)^{\mathrm{t}}(A\odot B)=A^{\mathrm{t}}A\ast B^{\mathrm{t}}B$.

Using block matrices and \cite{2matrices}, we can easily get the following result.
%
\begin{lemma}\label{ss}
For matrices $S_{i}\in \mathbb{R}^{\alpha_{i}\times \beta_{i}}$ and $P_{i}\in\mathbb{R}^{\beta_{i}\times r}$, we have that
\be\label{sf0}
(S_1\otimes S_{2}\otimes\cdots\otimes S_{N})(P_{1}\odot P_{2}\odot\cdots\odot P_{N})=(S_{1}P_{1})\odot (S_{2}P_{2})\odot\cdots \odot(S_{N}P_{N}).
\ee
\end{lemma}

\noindent
%


Recall the realignment of matrix \cite{realignment,realig}:
for any $m\times n$ matrix $Y=(y_{ij})$, $vec(Y)$ is the column vector
\begin{equation}\label{vec}
\emph{vec}(Y)=[y_{11},\cdots,y_{m1},y_{12}\cdots,y_{m2},\cdots,y_{1n},\cdots,y_{mn}]^{\mathrm{t}}.
\end{equation}

Similarly, let $Z$ be an $m\times m$ block matrix with each block of size $n\times n$, the realigned matrix $R(Z)$ is defined by
\begin{equation}\label{Rz}
R(Z)=[vec(Z_{11}),\cdots,vec(Z_{m1}),\cdots,vec(Z_{1m}),\cdots,vec(Z_{mm})]^{\mathrm{t}}.
\end{equation}

By \cite{rpm,rpms}, a necessary and sufficient condition for a matrix
being a tensor product of invertible matrices is the following result:
\begin{lemma}\label{3}\cite{rpms} For a $(d_1d_2\cdots d_N)\times (d_1d_2\cdots d_N$) invertible matrix $M$, there exist $d_i\times d_i$ invertible matrices $m_i(i=1,2,\cdots,N)$, such that $M=m_1\otimes m_2\otimes\cdots\otimes m_N$ if and only if rank $(R(M_{i|\widehat{i}})=1)$ for all $i$.
\end{lemma}

Here, for the matrix $M$ given in the lemma, 
$M_{i|\widehat{i}}$ denotes the $d_{i}\times d_{i}$ block matrix with each block of size $(d_1d_2\cdots d_{i-1}d_{i+1}\cdots d_N)\times (d_1d_2\cdots d_{i-1}d_{i+1}\cdots d_N)$, i.e., $M_{i|\widehat{i}}$ is a bipartite partitioned matrix of $M$.
In particular, when $N=2$, the necessary and sufficient condition to detect $M=m_{1}\otimes m_{2}$ is just that $rank(R(M))=1$.

\section{Criteria of SLOCC and LU equivalences for generic states}
In the first part, we consider the statistic local operations and classical communications (SLOCC) equivalence of two states.

Let $|\varphi\rangle$ and $|\psi\rangle$ be two pure states in Hilbert space $\mathcal{H}=\mathcal{H}_1\otimes\mathcal{H}_2\otimes\cdots\mathcal{H}_N$ with dimension $\prod^{N}_{i=1}d_{i}$, and $d_{i}=\mathrm{dim}\mathcal{H}_i,i=1,2,\cdots,N$. They are called SLOCC equivalent if 
\begin{equation}\label{slpure}
|\varphi\rangle=M_1\otimes M_2\otimes\cdots\otimes M_{N}|\psi\rangle
\end{equation}
for invertible local operators (ILOs) $M_i$ in $GL(d_i,\mathbb{C})$, $1\leq i\leq N$.

Two mixed states $\rho$ and $\rho'$ are said to be SLOCC equivalent if and only if there exist $M_i\in GL(d_i,\mathbb{C}),i=1,\cdots,N$ such that
\begin{equation}\label{slmix}
\rho'=(M_1\otimes M_2\otimes\cdots\otimes M_{N})\rho(M_1\otimes M_2\otimes\cdots\otimes M_{N})^{\dag}.
\end{equation}

Now we study pure states based on their coefficient tensors.
For any $N$ partite pure state
\begin{equation}\label{purefj}
|\varphi\rangle=\sum_{j_1,j_2\cdots,j_{N}=1}^{d_1,d_2,\cdots,d_{N}}\alpha_{j_1j_2\cdots j_N}|j_1j_2\cdots j_N\rangle
\end{equation}with
$\sum_{j_1,j_2\cdots,j_{N}=1}^{d_1,d_2,\cdots,d_{N}}|\alpha_{j_1j_2\cdots j_N}|^{2}=1$ and
$\{|j_t\rangle_{j_t=1}^{d_t}\}$ 
are orthonormal bases of the Hilbert space $\mathcal{H}_{t}$ $( 1\leq t\leq N)$. We associate the $N$th-order coefficient tensor $\mathcal{X}=(\alpha_{j_1j_2\cdots j_N})$.

For example, for a $3$-qutrit pure state $|\psi\rangle=\sum_{i_1,i_2,i_3=1}^{3}x_{i_1i_2i_3}|i_1i_2i_3\rangle$, the frontal slices
are $X_{1}=\left(\begin{smallmatrix}
     x_{111} & x_{121} & x_{131}\\
     x_{211} & x_{221} & x_{231}\\
     x_{311} & x_{321} & x_{331}
\end{smallmatrix}\right)$, $X_{2}=\left(\begin{smallmatrix}
     x_{112} & x_{122} & x_{132}\\
     x_{212} & x_{222} & x_{232}\\
     x_{312} & x_{322} & x_{332}
\end{smallmatrix}\right)$, $X_{3}=\left(\begin{smallmatrix}
     x_{113} & x_{123} & x_{133}\\
     x_{213} & x_{223} & x_{233}\\
     x_{313} & x_{323} & x_{333}
\end{smallmatrix}\right)$.  Then the three mode-n matricizations are
\begin{equation}\label{modex}
\begin{array}{ll}
X_{(1)}=\scriptsize{\left(\begin{matrix}
     x_{111} & x_{121} & x_{131} & x_{112} & x_{122} & x_{132} & x_{113} & x_{123} & x_{133}\\
     x_{211} & x_{221} & x_{231} & x_{212} & x_{222} & x_{232} & x_{213} & x_{223} & x_{233}\\
     x_{311} & x_{321} & x_{331} & x_{312} & x_{322} & x_{332} & x_{313} & x_{323} & x_{333}
\end{matrix}\right)}, \vspace{0.5em}\\
X_{(2)}=\scriptsize{\left(\begin{matrix}
     x_{111} & x_{211} & x_{311} & x_{112} & x_{212} & x_{312} & x_{113} & x_{213} & x_{313}\\
     x_{121} & x_{221} & x_{321} & x_{122} & x_{222} & x_{322} & x_{123} & x_{223} & x_{323}\\
     x_{131} & x_{231} & x_{331} & x_{132} & x_{232} & x_{332} & x_{133} & x_{233} & x_{333}
\end{matrix}\right)},  \vspace{0.5em}\\
X_{(3)}=\scriptsize{\left(\begin{matrix}
     x_{111} & x_{211} & x_{311} & x_{121} & x_{221} & x_{321} & x_{131} & x_{231} & x_{331}\\
     x_{112} & x_{212} & x_{312} & x_{122} & x_{222} & x_{322} & x_{132} & x_{232} & x_{332}\\
     x_{113} & x_{213} & x_{313} & x_{123} & x_{223} & x_{323} & x_{133} & x_{233} & x_{333}\\
\end{matrix}\right)}.
\end{array}
\end{equation}
\vspace{-1em}
\begin{theorem}\label{pureth1}
Two $N$-partite pure states $|\varphi\rangle$ and $|\psi\rangle$ on $\mathcal{H}$ are SLOCC equivalent if and only if their coefficient tensors $\mathcal{X}$ and $\mathcal{Y}$
satisfy
\be\label{puretheo1}
\mathcal{Y}=(M_1\otimes M_2\otimes\cdots\otimes M_N)\mathcal{X},\ M_{i}\in GL(d_i).
\ee
\end{theorem}

\noindent
{\bf{Proof:}} Let $|\varphi\rangle$ be a pure state written as \eqref{purefj}. Suppose 
there are invertible matrices $M_{i}(i=1,2,\cdots,N)$ such that $|\psi\rangle=(M_1\otimes M_2\otimes\cdots\otimes M_{N})|\varphi\rangle$, then
\begin{eqnarray}
\begin{array}{ll}
|\psi\rangle&=(M_{1}\otimes M_{2}\otimes\cdots\otimes M_{N})|\varphi\rangle\\
&=\sum\limits_{j_1,j_2\cdots,j_{N}=1}^{d_1,d_2,\cdots,d_{N}}\alpha_{j_1j_2\cdots j_N}(M_{1}|j_1\rangle)\otimes (M_{2}|j_2\rangle)\otimes\cdots\otimes (M_{N}|j_N\rangle) \\
&=\sum\limits_{j_1,j_2\cdots,j_{N}=1}^{d_1,d_2,\cdots,d_{N}}\alpha_{j_1j_2\cdots j_N}\scriptsize\begin{pmatrix}
     m^{(1)}_{1j_1}\\
     \vdots\\
     m^{(1)}_{d_{1}j_1}
\end{pmatrix}\otimes\scriptsize\begin{pmatrix}
     m^{(2)}_{1j_2}\\
     \vdots\\
     m^{(2)}_{d_{2}j_2}
\end{pmatrix}\otimes\cdots\otimes\scriptsize\begin{pmatrix}
     m^{(N)}_{1j_N}\\
     \vdots\\
     m^{(N)}_{d_{N}j_N}
\end{pmatrix} \\
&=(\beta_{11\cdots 11}, \cdots, \beta_{11\cdots 1d_N}, \cdots, \beta_{d_1d_2\cdots d_{N-1}1}, \cdots, \beta_{d_1d_2\cdots d_{N-1}d_N})^{\mathrm{t}},
\end{array}
\end{eqnarray}
where $\beta_{j_1j_2\cdots j_{N}}=\sum^{d_1,d_2\cdots,d_{N}}_{i_1,i_2\cdots,i_{N}=1}m^{(1)}_{j_1i_1}
m^{(2)}_{j_2i_2}\cdots m^{(N)}_{j_Ni_N}\alpha_{i_1i_2\cdots i_{N}}$
 and $\sum_{j_1,j_2\cdots,j_{N}=1}^{d_1,d_2,\cdots,d_{N}}|\beta_{j_1j_2\cdots j_N}|^{2}=1$, which follows from the norm condition of $|\psi\rangle$.


Therefore $\left((M_1\otimes \cdots \otimes M_N)\mathcal X\right)_{j_1j_2\dots j_N}=\sum\limits_{i_1,i_2,\dots, i_N} m^{(1)}_{j_1i_1}m^{(2)}_{j_2i_2}\cdots m^{(N)}_{j_Ni_N}\alpha_{i_1i_2\dots i_N}=\beta_{j_1j_2\dots j_N}=\left(\mathcal{Y}\right)_{j_1j_2\dots j_N}$ by (\ref{tens}),
i.e. $\mathcal{Y}=(M_1\otimes M_2\otimes\cdots\otimes M_N)\mathcal{X}$.

Conversely suppose there exist the coefficient tensors $\mathcal{X}$ and $\mathcal{Y}$ of pure states $|\varphi\rangle$ and $|\psi\rangle$ such that  $\mathcal{Y}=(M_1\otimes M_2\otimes\cdots\otimes M_N)\mathcal{X}$ for invertible matrices $M_i(i=1,2,\cdots,N)$, i.e. $\beta_{j_1j_2\cdots j_{N}}=\sum^{d_1,d_2\cdots,d_{N}}_{i_1,i_2\cdots,i_{N}=1}m^{(1)}_{j_1i_1}
m^{(2)}_{j_2i_2}\cdots m^{(N)}_{j_Ni_N}\alpha_{i_1i_2\cdots i_{N}}$. Therefore, Eq.(\ref{slpure}) holds true, so $|\varphi\rangle$ and $\psi$ are SLOCC equivalent.
$\hfill\Box$

\begin{lemma}\label{mm} Let $\mathcal{X}$ and $\mathcal{Y}$ be coefficient tensors of SLOCC equivalent pure states
$|\varphi\rangle$ and $|\psi\rangle$ respectively, then
Rank$(\mathcal{X})$=Rank$(\mathcal{Y})$.
\end{lemma}

\noindent
{\bf{Proof:}} Let $R$ be the rank of the coefficient tensor $\mathcal{X}$ of $|\varphi\rangle$. The CP decomposition says that
$\mathcal{X}$ can be written as
\begin{equation}\label{NcX}
\mathcal{X}=\sum^{R}_{r=1}\mathbf{a}^{(1)}_{r}\circ\mathbf{a}^{(2)}_{r}\circ\cdots\circ\mathbf{a}^{(N)}_{r}=[\![{A_{1}},{A_{2}},\cdots,{A_{N}}]\!].
\end{equation}
If $|\varphi\rangle$   is SLOCC equivalent to $|\psi\rangle$, then Theorem \ref{pureth1} implies that
 $\mathcal{Y}=(M_1\otimes M_2\otimes\cdots\otimes M_N)\mathcal{X}=\sum^{R}_{r=1}(M_1\mathbf{a}^{(1)}_{r})\circ \cdots\circ (M_{N}\mathbf{a}^{(N)}_{r})$
for invertible local operators (ILOs) $M_i\in GL(d_i,\mathbb{C})$ for each $i$. We obtain Rank($\mathcal{Y}$)$\leq R$ by the definition of tensor rank.
Switching $|\varphi\rangle$ and $|\psi\rangle$, we get that Rank$(\mathcal{X})$=Rank$(\mathcal{Y})$.
$\hfill\Box$

\begin{theorem}\label{pureth2}
Two $K$-partite pure states $|\varphi\rangle$ and $|\psi\rangle$ are SLOCC equivalent if and only if the mode-$n$ matricizations $X_{(i)}$ and $Y_{(i)}$ of their coefficient tensors $\mathcal{X}$ and $\mathcal{Y}$ satisfy that
\be\label{puretheo}
Y_{(i)}=M_{i}X_{(i)}(M_N\otimes\cdots\otimes M_{i+1}\otimes M_{i-1}\otimes\cdots\otimes M_1)^{\mathrm{t}},
\ee
where $M_{i}\in GL(d_i) (i=1,2,\cdots,N)$.
\end{theorem}

\noindent
{\bf{Proof:}} It follows from
SLOCC equivalence and the CP decomposition of the coefficient tensor that there exist $d_i\times d_i$ invertible matrices $M_{i}(i=1,\cdots,N)$ such that two coefficient tensors $\mathcal{X}$ and $\mathcal{Y}$ have the following form
\begin{equation}\label{NNcY}
\mathcal{Y}=\sum^{R}_{r=1}\mathbf{b}^{(1)}_{r}\circ\mathbf{b}^{(2)}_{r}\circ\cdots\circ\mathbf{b}^{(N)}_{r}=\sum^{R}_{r=1}(M_1\mathbf{a}^{(1)}_{r})\circ \cdots\circ (M_{N}\mathbf{a}^{(N)}_{r})=(M_1\otimes M_2\otimes\cdots\otimes M_N)\mathcal{X}.
\end{equation}
So the factor matrices $A_{i}$ and $B_{i}$ obey
$B_{i}=(\mathbf{b}^{(i)}_1 \ \mathbf{b}^{(i)}_2 \ \cdots \ \mathbf{b}^{(i)}_{N})=M_iA_i$.

By (\ref{nmatricization}) and (\ref{ss}), we also have
\begin{equation*}
\begin{array}{ll}
Y_{(i)}&=B_{i}(B_{N}\odot\cdots\odot B_{i+1}\odot B_{i-1}\odot \cdots\odot B_{1})^\mathrm{t}\\
&=(M_{i}A_{i})((M_{N}\otimes\cdots\otimes M_{i+1}\otimes M_{i-1}\cdots\otimes M_{1})(A_{N}\odot\cdots\odot A_{i+1}\odot A_{i-1}\odot\cdots\odot A_{1}))^\mathrm{t}\\
&=M_{i}(A_{i}(A_{N}\odot\cdots\odot A_{i+1}\odot A_{i-1}\odot\cdots\odot A_{1})^\mathrm{t})(M_{N}\otimes\cdots\otimes M_{i+1}\otimes M_{i-1}\cdots\otimes M_{1})^\mathrm{t}\\
&=M_{i}X_{(i)}(M_N\otimes\cdots\otimes M_{i+1}\otimes M_{i-1}\otimes\cdots\otimes M_1)^{\mathrm{t}},
\end{array}
\end{equation*}
where $i=1,\cdots,N$.

Conversely, if there are invertible matrices $M_{i}$, such that $Y_{(i)}=M_{i}X_{(i)}(M_N\otimes\cdots\otimes M_{i+1}\otimes M_{i-1}\otimes\cdots\otimes M_1)^{\mathrm{t}}$. Suppose $B_{i}=M_{i}A_{i}$, we 
get that $\mathcal{Y}=(M_1\otimes M_2\otimes\cdots\otimes M_N)\mathcal{X}$.  The converse direction can be shown similarly. 
$\hfill\Box$

The result implies that the rank of any mode-$n$ matricization $X_{(i)}(i=1,2,\cdots,N)$ of coefficient tensor $\mathcal{X}$ of an $N$-partite pure state is invariant under SLOCC. When $N=2$, the rank invariance of $X_{(i)}$ is a necessary and sufficient condition for SLOCC equivalence of two states.


\begin{theorem}\label{pureth3}
Two $N$-partite pure states $|\varphi\rangle$ and $|\psi\rangle$ are SLOCC equivalent if and only if the
mode-$n$ matricizations $X_{(i)}$ and $Y_{(i)}$ of their coefficient tensors $\mathcal{X}$ and $\mathcal{Y}$ 
satisfy the condition: there are ${d_i\times d_i}$ invertible matrix $P_{i}$
and $(d_1\cdots d_{i-1}d_{i+1}\cdots d_{N})\times (d_1\cdots d_{i-1}d_{i+1}\cdots d_{N})$ invertible matrix $Q_{i}$
such that
\be\label{puretheoo}
Y_{(i)}=P_{i}X_{(i)}Q_{i}^{\mathrm{t}},
\ee
and
\be\label{puretheoo1}
rank(R(Q_{i})_{j|\widehat{j}})=1,
\ee
where $i,j=1,\cdots,N$ and $j\neq i$.
\end{theorem}

\noindent
{\bf{Proof:}} Combining with Theorem \ref{pureth2} and Lemma \ref{3}, we can easily see if
two $K$-partite pure states $|\varphi\rangle$ and $|\psi\rangle$ are SLOCC equivalent,
then $Y_{(i)}=M_{i}X_{(i)}(M_N\otimes\cdots\otimes M_{i+1}\otimes M_{i-1}\otimes\cdots\otimes M_1)^{\mathrm{t}}$, where $M_{i}(i=1,2,\cdots,N)$ are $d_i\times d_i$ invertible matrices, i.e. there are invertible matrices $P_i=M_{i}$ and $Q_{i}=M_N\otimes\cdots\otimes M_{i+1}\otimes M_{i-1}\otimes\cdots\otimes M_1$, such that $Y_{(i)}=P_{i}X_{(i)}Q_{i}^{\mathrm{t}}$ and $rank(R(Q_{i})_{j|\widehat{j}})=1$, $j=1,2,\cdots,i-1,i+1,\cdots,N$.

Conversely, suppose there exist mode-$n$ matricizations $X_{(i)}$ and $Y_{(i)}$ of coefficient tensors $\mathcal{X}$ and $\mathcal{Y}$ satisfying the condition (\ref{puretheoo}) and (\ref{puretheoo1}) of the Theorem.
Then, there are invertible matrices $M_{i}(i=1,2,\cdots,N)$ such that (\ref{puretheo}) holds true by Lemma \ref{3}. 
By Theorem \ref{pureth2}, $|\varphi\rangle$ and $|\psi\rangle$ are SLOCC equivalent.
$\hfill\Box$

In practice, for the mode-$n$ matricizations $X_{(i)}$ and $Y_{(i)}$ of the coefficient tensors $\mathcal{X}$ and $\mathcal{Y}$,
we first get two equations $X_{(i)}=U^{(i)}_{1}\Sigma_{i}V^{(i)}_{1}$ and $Y_{(i)}=U^{(i)}_{2}\Lambda_{i}V^{(i)}_{2}$ by the singular value decomposition. Then
$\Sigma_{i}=D_{i}\Lambda_{i}W_{i}$, where $\Lambda_{i}=diag(\lambda_1,\ \lambda_2,\ \cdots, \ \lambda_r,\ 0,\ \cdots, \ 0)$, $\Sigma_{i}=diag(\sigma_1,\ \sigma_2,\ \cdots, \ \sigma_r,\ 0,\ \cdots,\ 0)$, $d_i\times d_i$ invertible matrix $D_i=diag(\sqrt{\frac{\sigma_1}{\lambda_1}},\ \sqrt{\frac{\sigma_2}{\lambda_2}},\ \cdots,\ \sqrt{\frac{\sigma_r}{\lambda_r}},\ 1,\ \cdots,\ 1)$, and $(d_1\cdots d_{i-1}d_{i+1}\cdots d_{N})\times (d_1\cdots d_{i-1}d_{i+1}\cdots d_{N})$ invertible matrix $W_i=diag(\sqrt{\frac{\sigma_1}{\lambda_1}},\ \sqrt{\frac{\sigma_2}{\lambda_2}},\ \cdots,\ \sqrt{\frac{\sigma_r}{\lambda_r}},\ 1,\ \cdots,\ 1)$. Combining these, we can find the invertible matrices $P_{i}$ and $Q_{i}$ of (\ref{puretheoo}). Finally, we just check (\ref{puretheoo1}) to see whether the two pure states are SLOCC equivalent.

In fact, with the above method, as long as one pair of $X_{(i)}$ and $Y_{(i)}$ satisfies the condition of Theorem \ref{pureth3}, we can deduce that two states are SLOCC equivalent without verifying all $i$.

{\bf Example 1:} Consider two 3-qubit pure states:
\begin{equation}\label{GHZ}
|GHZ\rangle=\frac{1}{\sqrt{2}}(|000\rangle+|111\rangle),
\end{equation}
and
\begin{equation}\label{state}
|\psi\rangle=\frac{1}{2}(|000\rangle+|001\rangle+|110\rangle-|111\rangle).
\end{equation}

The third-order coefficient tensor $\mathcal{X}=(x_{i_1i_2i_3})$ of state $|GHZ\rangle$ 
is given by $x_{111}=x_{222}=\frac{1}{\sqrt{2}}$, and the other elements are equal to 0.
By (\ref{nmatricization}), 
the three mode-$n$ unfoldings are
\begin{equation}\label{X123}
X_{(1)}=X_{(2)}=X_{(3)}=
\left(\begin{smallmatrix}
     \frac{1}{\sqrt{2}} & 0 & 0 & 0\\
     0 & 0 & 0 & \frac{1}{\sqrt{2}}
\end{smallmatrix}\right).
\end{equation}

Similarly, the coefficient tensor of state $|\psi\rangle$ is $\mathcal{X}'=(x'_{i_1i_2i_3})$ with
$x'_{111}=x'_{112}=x'_{221}=\frac{1}{2}$, $x'_{222}=-\frac{1}{2}$, and $x'_{121}=x'_{122}=x'_{211}=x'_{212}=0$. Therefore,
\begin{equation}\label{X'123}
X'_{(1)}=X'_{(2)}=
\left(\begin{smallmatrix}
     \frac{1}{2} & 0 & \frac{1}{2} & 0\\
     0 & \frac{1}{2} & 0 & -\frac{1}{2}
\end{smallmatrix}\right) \
\mathrm{and} \
X'_{(3)}=
\left(\begin{smallmatrix}
     \frac{1}{2} & 0 & 0 & \frac{1}{2}\\
     \frac{1}{2} & 0 & 0 & -\frac{1}{2}
\end{smallmatrix}\right).
\end{equation}

We consider the matricization form $X_{(1)}$ and $X'_{(1)}$ of tensors $\mathcal{X}$ and $\mathcal{X}'$, respectively. 
Using the singular value decomposition, 
we get
\begin{equation}\label{X1}
X_{(1)}=
\left(\begin{smallmatrix}
     1 & 0\\
     0 & 1
\end{smallmatrix}\right)
\left(\begin{smallmatrix}
     \frac{1}{\sqrt{2}} & 0 & 0 & 0\\
     0 & \frac{1}{\sqrt{2}} & 0 & 0
\end{smallmatrix}\right)
\left(\begin{smallmatrix}
     1 & 0 & 0 & 0\\
     0 & 0 & 0 & -1 \\
     0 & 0 & 1 & 0 \\
     0 & 1 & 0 & 0
\end{smallmatrix}\right)^{\mathrm{t}}
=U_1S_1V^{\mathrm{t}}_1
\end{equation}
and
\begin{equation}\label{X'1}
X'_{(1)}=
\left(\begin{smallmatrix}
     1 & 0\\
     0 & 1
\end{smallmatrix}\right)
\left(\begin{smallmatrix}
     \frac{1}{\sqrt{2}} & 0 & 0 & 0\\
     0 & \frac{1}{\sqrt{2}} & 0 & 0
\end{smallmatrix}\right)
\left(\begin{smallmatrix}
     \frac{1}{\sqrt{2}} & 0 & -\frac{1}{\sqrt{2}} & 0\\
     0 & \frac{1}{\sqrt{2}} & 0 & \frac{1}{\sqrt{2}} \\
     \frac{1}{\sqrt{2}} & 0 & \frac{1}{\sqrt{2}} & 0 \\
     0 & -\frac{1}{\sqrt{2}} & 0 & \frac{1}{\sqrt{2}}
\end{smallmatrix}\right)^{\mathrm{t}}
=U'_1S'_1V'^{\mathrm{t}}_1.
\end{equation}
It follows from (\ref{X1}) and (\ref{X'1}) that
\begin{equation}\label{X1X'1}
X'_{(1)}=U'_1S'_1V'^{\mathrm{t}}_1=U_1S_1V^{\mathrm{t}}_1V_1V'^{\mathrm{t}}_1
=\mathrm{I}_{2\times2}X_{(1)}(V_1V'^{\mathrm{t}}_1)
\end{equation}
Let $P_1=\mathrm{I}_{2\times2}$ and $Q_1=V'_1V^{\mathrm{t}}_1=\left(\begin{smallmatrix}
     \frac{1}{\sqrt{2}} & 0 & -\frac{1}{\sqrt{2}} & 0\\
     0 & -\frac{1}{\sqrt{2}} & 0 & \frac{1}{\sqrt{2}} \\
     \frac{1}{\sqrt{2}} & 0 & \frac{1}{\sqrt{2}} & 0 \\
     0 & -\frac{1}{\sqrt{2}} & 0 & -\frac{1}{\sqrt{2}}
\end{smallmatrix}\right)$, 
then $rank(R(Q_1))=1$ and $Q_1=M_3\otimes M_2$.
Therefore, for
$M_1=\left(\begin{smallmatrix}
     1 & 0\\
     0 & 1\\
\end{smallmatrix}\right)$, $M_2=\left(\begin{smallmatrix}
     1 & 0\\
     0 & -1\\
\end{smallmatrix}\right)$, and $M_3=\left(\begin{smallmatrix}
     \frac{1}{\sqrt{2}} & -\frac{1}{\sqrt{2}}\\
     \frac{1}{\sqrt{2}} & \frac{1}{\sqrt{2}}\\
\end{smallmatrix}\right)$, we get $X'_{(1)}=M_1X_{(1)}(M_3\otimes M_2)^{\mathrm{t}}$.
By Theorem \ref{pureth3}, we know the state $|GHZ\rangle$ is SLOCC equivalent to $|\psi\rangle$.

Next we study two mixed states $\rho$ and $\rho'$.

If they are equivalent under SLOCC, by (\ref{slmix}) we have that 
$\rho'=(M_1\otimes M_2\otimes\cdots\otimes M_{N})\rho(M_1\otimes M_2\otimes\cdots\otimes M_{N})^{\dag}$, where $M_{k}\in GL(d_k,\mathbb{C})$ for $k=1,\cdots,N$.
Let $\lambda^{(k)}_{i}$ be the Gell-Mann basis elements, suppose
\begin{equation}\label{Nmi}
   M_{k}\lambda^{(k)}_{i-1}M_{k}^{\dagger}=\sum^{d_k^2}_{j=1}L^{(k)}_{ji}\lambda^{(k)}_{j-1},
\end{equation}
where $L_k=(L^{(k)}_{ij})$ is a $d^2_k\times d^2_k$ invertible matrix and $\lambda^{(k)}_{0}=\mathrm{I}_{d_k}$. 

Then on a $d_1\times\cdots\times d_N$ dimensional Hilbert space $H^{d_1}_1\otimes H^{d_2}_2\otimes\cdots\otimes H^{d_N}_{N}$, a multi-qudit state $\rho$ can be expressed as
\begin{equation}\label{mrho}
\rho=\sum^{d^2_1}_{i_1=1}\sum^{d^2_2}_{i_2=1}\cdots\sum^{d^2_N}_{i_N=1}x_{i_1i_2\cdots i_N}\lambda^{(1)}_{i_1-1}\otimes\lambda^{(2)}_{i_2-1}\otimes\cdots\otimes\lambda^{(N)}_{i_N-1},
\end{equation}
for $
x_{i_1i_2\cdots i_N}=2^{M-N}(d_{k_0}d_{k_1}\cdots d_{k_M})^{-1}\mathrm{Tr}(\rho\lambda^{(1)}_{i_1}\otimes\lambda^{(2)}_{i_2}\otimes\cdots\lambda^{(N)}_{i_N})$,
where there are exactly $M$-superscripts equal to $0$: $i_{k_{1}}=i_{k_{2}}=\cdots=i_{k_{M}}=0$, and $d_{k_0}=1$, $M\in \{0,1,\cdots,N\}$, $k_1,k_2,\cdots,k_M\in \{1,2,\cdots,N\}$, $k_{l}(l=1,\cdots,M)$ are not equal to each other.

Like the pure state, we use $\mathcal{X}=(x_{i_1i_2\cdots i_N})$ to denote
an $N$-order real coefficient tensor of multipartite state $\rho$.
Then the coefficient tensors $\mathcal{X}$ and $\mathcal{X}'$ of two SLOCC equivalent states satisfy the equation $(L_1\otimes\cdots\otimes L_N)\mathcal{X}=\mathcal{X}'$. 
By the CP decomposition, 
$\mathcal{X}'=(L_1\otimes\cdots\otimes L_N)\mathcal{X}=(L_1\otimes\cdots\otimes L_N)\sum^{R}_{r=1}\mathbf{a}^{(1)}_{r}\circ\cdots\circ\mathbf{a}^{(N)}_{r}=\sum^{R}_{r=1}L_1\mathbf{a}^{(1)}_{r}\circ\cdots\circ L_N\mathbf{a}^{(N)}_{r}
=[\![L_1{A_{1}},\cdots,{L_NA_{N}}]\!]$. 
The following is a necessary condition of two SLOCC equivalent states.

\begin{lemma}\label{mixle}
Suppose that the $N$-partite mixed states $\rho$ and $\rho'$ over $\otimes_{i=1}^NH_i$ are SLOCC equivalent, where $dim(H_i)=d_i$, then there are $d^2_i\times d^2_i$ invertible matrices $L_{i}(i=1,2,\cdots,N)$ such that the mode-$n$ matricizations $X_{(i)}$ and $X'_{(i)}$ of their coefficient tensors $\mathcal{X}$ and $\mathcal{X}'$ satisfy
\be\label{purethslmieo}
X'_{(i)}=L_{i}X_{(i)}(L_N\otimes\cdots\otimes L_{i+1}\otimes L_{i-1}\otimes\cdots\otimes L_1)^{\mathrm{t}}.
\ee
\end{lemma}

In practice we can use the
singular value decomposition and Lemma \ref{mixle} to get hold of the congruent transformation. %
The following result follows easily from Lemma \ref{mixle}.

\begin{theorem}\label{mixth} Let $H=\otimes_{i=1}^NH_i$ be the Hilbert space with $dim(H_i)=d_i$.
If two $N$-partite mixed quantum states $\rho$ and $\rho'$ over $H$ are SLOCC equivalent, then there are ${d^2_i\times d^2_i}$ unitary matrices $P^{(1)}_{i}$, $P^{(2)}_{i}$, invertible diagonal matrices $S_1, S_2$ and ${d^2_1\cdots d^2_{i-1}d^2_{i+1}\cdots d^2_{N}\times d^2_1\cdots d^2_{i-1}d^2_{i+1}\cdots d^2_{N}}$ unitary matrices $Q^{(1)}_{i}$, $Q^{(2)}_{i}$ such that
\be\label{puretheslmoo}
X'_{(i)}=P^{(1)}_{i}S_1(P^{(2)}_{i})^{\dag}X_{(i)}(Q^{(2)}_{i})^{\dag}S_2Q^{(1)}_{i} \ \ and \ \
rank(R((Q^{(2)}_{i})^{\dag}S_2Q^{(1)}_{i})_{j|\widehat{j}})=1,
\ee
where $i,j=1,2,\cdots,N$ and $i\neq j$. $X_{(i)}$ and $X'_{(i)}$ are the mode-$n$ matricizations of the coefficient tensors $\mathcal{X}$ and $\mathcal{X}'$ of two states, respectively.
\end{theorem}


In the second part, we study LU equivalence of quantum states with the same methodology.
Similar to SLOCC equivalence, the quantum state $|\varphi\rangle$ is local unitary (LU) equivalent to $|\psi\rangle$ if
$|\varphi\rangle=U_1\otimes U_2\otimes\cdots\otimes U_{N}|\psi\rangle$, when $U_i(i=1,\cdots,N)$ are unitary operators. The following is clear.

\begin{theorem}\label{purethlu}
Two $N$-partite pure states $|\varphi\rangle$ and $|\psi\rangle$ over $H=\otimes_{i=1}^NH_i$ with $dim(H_i)=d_i$ are LU equivalent if and only if their coefficient tensors $\mathcal{X}$ and $\mathcal{X}'$ satisfy
\be\label{puretheolu1}
\mathcal{X}'=(U_1\otimes U_2\otimes\cdots\otimes U_N)\mathcal{X}.
\ee
for $d_i\times d_i$ unitary matrices $U_{i}(i=1,2,\cdots,N)$.
\end{theorem}

Similarly, the following results are obtained (by appropriately changing invertible matrix to unitary matrix).

Two $N$-partite pure states $|\varphi\rangle$ and $|\psi\rangle$ over $H=\otimes_{i=1}^NH_i$ with $dim(H_i)=d_i$ are LU equivalent if and only if their mode-$n$ matricizations $X_{(i)}$ and $X'_{(i)}$ of coefficient tensors $\mathcal{X}$ and $\mathcal{X}'$ satisfy the equations $X_{(i)}=U_{i}X'_{(i)}(U_N\otimes\cdots\otimes U_{i+1}\otimes U_{i-1}\otimes\cdots\otimes U_1)^{\mathrm{t}}$, where $U_i$ is $d_i\times d_i$ a unitary matrix for $i=1,\cdots,N$.
In other words, we can also obtain the following  necessary and sufficient condition.

\begin{theorem}\label{purethlu3} Let $H=\otimes_{i=1}^NH_i$ with $dim(H_i)=d_i$ be the Hilbert space.
An $N$-partite pure state $|\varphi\rangle$ is LU equivalent to an $N$-partite pure state $|\psi\rangle$ if and only if  their mode-$n$ matricizations $X_{(i)}$ and $X'_{(i)}$ of coefficient tensors $\mathcal{X}$ and $\mathcal{X}'$ obey the following equations: for $i\neq j=1,2,\cdots,N$
\be\label{puretheluoo}
X'_{(i)}=U_{i}X_{(i)}Q_{i}^{\mathrm{t}},
\ee
for unitary matrices $U_{i}\in SU(d_i), Q_{i}\in SU(d/d_i)$ and $ rank(R(Q_{i})_{j|\widehat{j}})=1$. Here $d=d_1\cdots d_N$.
\end{theorem}

Next for mixed states, we start with a necessary and sufficient condition of LU equivalence for 2-partite states based on the spectral decomposition and Lemma \ref{3} (cf. \cite{FeiJing}).

\begin{theorem}\label{mixthlu1}
Two $2$-partite mixed states $\rho$ and $\rho'$ over a $d_1\otimes d_2$-Hilbert space are LU equivalent if and only if 
there exists a ${d_1d_{2}\times d_1d_{2}}$ unitary matrix $P$ such that
\be\label{mix2theoo}
\rho'=P\rho P^{\dag},
\ee
and
\be\label{mix2theoo1}
rank(R(P))=1,
\ee
\end{theorem}


\noindent
{\bf{Proof:}} Suppose $2$-partite mixed states $\rho$ and $\rho'$ are LU equivalent, 
then there exist unitary matrices $U_1$, $U_2$ such that $\rho'=(U_1\otimes U_2)\rho(U_1\otimes U_2)^{\dag}$. It is clear that $\rho$ and $\rho'$ have the same eigenvalues. 
Write $U_1\otimes U_2=X\Lambda X^{\dag}$, where $X=[\mathbf{x}_1,\mathbf{x}_2,\cdots,\mathbf{x}_{d_1d_{2}}]$ and $\{\mathbf{x}_i\}(i=1,\cdots,d_1d_{2})$ are the normalized eigenvectors of the unitary matrix $U_1\otimes U_2$.
Take $P=X\Lambda X^{\dag}$, then $rank(R(P)=rank(R(U_1\otimes U_2))=1$ by Lemma \ref{3}.

On the other hand, suppose there exists $d_1d_2\times d_1d_2$ unitary matrix $P$ satisfying the conditions (\ref{mix2theoo}) and (\ref{mix2theoo1}) above.
By Lemma \ref{3} and $rank(R(P))=1$, there exist two invertible matrices $Q_1$ and $Q_2$ such that $P=Q_1\otimes Q_2$. Since $P$ is unitary,
$(Q_1Q^{\dag}_1)\otimes(Q_2Q^{\dag}_2)=I$.
Thus $Q_1Q_1^{\dag}$ and $Q_2Q_2^{\dag}$ are (real) scalar matrices, say $Q_1Q_1^{\dag}=aI$ and $Q_2Q_2^{\dag}=a^{-1}I$. Take the unitary matrices $U_1={(1/\sqrt{a})}Q_1$ and $U_2={\sqrt{a}}Q_2$, then $\rho'=(U_1\otimes U_2)\rho(U_1\otimes U_2)^{\dag}$.
$\hfill\Box$

Using induction, we can easily get the LU equivalence condition for N-partite mixed states.

\begin{theorem}\label{T:mixN} Two $N$-partite mixed states $\rho$ and $\rho'$ over
the Hilbert space $H=\otimes_{i=1}^NH_i$ with $dim(H_i)=d_i$ are LU equivalent if and only if there exists a $({d_1\cdots d_{N})\times (d_1\cdots d_{N}})$ unitary matrix $P$ such that
\be\label{mixNtheoon}
\rho'=P\rho P^{\dag},
\ee
and
\be\label{mixNtheoon1}
rank(R(P)_{j|\widehat{j}})=1
\ee
for each $j=1,2,\cdots,N$.
\end{theorem}

Note that \eqref{mixNtheoon} implies that if two mixed states are LU equivalent, then they have same rank and eigenvalues.
To find the unitary matrix $P$, we can consider the spectral decomposition of two density matrices with same eigenvalues, i.e., for two LU equivalent states $\rho$ and $\rho'$, there are unitary matrices $P_1$, $P_2$ and real diagonal invertible matrix $\Lambda$ such that $\rho'=P_1\Lambda P_1^{\dag}$ and $\rho=P_2\Lambda P_2^{\dag}$. Then $\rho'=P_1P_2^{\dag}\rho P_2P_1^{\dag}=P\rho P^{\dag}$, where $P=P_2P_1^{\dag}$.  Eq. \eqref{mixNtheoon1} then ensures that $P$ obeys the locality condition.

{\bf Example 2:} Consider the following two 3-qubit mixed states \cite{rpms}:
\begin{equation}\label{rhok}
\rho=\frac{1}{K}
\left(\begin{smallmatrix}
     1 & 0 & 0 & 0 & 0 & 0 & 0 & 1\\
     0 & a & 0 & 0 & 0 & 0 & 0 & 0\\
     0 & 0 & b & 0 & 0 & 0 & 0 & 0\\
     0 & 0 & 0 & c & 0 & 0 & 0 & 0\\
     0 & 0 & 0 & 0 & \frac{1}{c} & 0 & 0 & 0\\
     0 & 0 & 0 & 0 & 0 & \frac{1}{b} & 0 & 0\\
     0 & 0 & 0 & 0 & 0 & 0 & \frac{1}{a} & 0\\
     1 & 0 & 0 & 0 & 0 & 0 & 0 & 1
\end{smallmatrix}\right),
\end{equation}
and
\begin{equation}\label{rho'k}
\rho'=\frac{1}{K}\left(\begin{smallmatrix}
     c & 0 & 0 & 0 & 0 & 0 & 0 & 0\\
     0 & b & 0 & 0 & 0 & 0 & 0 & 0\\
     0 & 0 & a & 0 & 0 & 0 & 0 & 0\\
     0 & 0 & 0 & 1 & -1 & 0 & 0 & 0\\
     0 & 0 & 0 & -1 & 1 & 0 & 0 & 0\\
     0 & 0 & 0 & 0 & 0 & \frac{1}{a} & 0 & 0\\
     0 & 0 & 0 & 0 & 0 & 0 & \frac{1}{b} & 0\\
      0& 0 & 0 & 0 & 0 & 0 & 0 & \frac{1}{c}
\end{smallmatrix}\right),
\end{equation}
where the normalization factor $K=2+a+b+c+\frac{1}{a}+\frac{1}{b}+\frac{1}{c}$. It is easy to see that both matrices are equivalent to
$\Lambda=\frac{1}{K}diag(\frac{1}{c},\frac{1}{b},\frac{1}{a},2,a,b,c,0)$.
To simplify the situation, we consider the case where $a\neq b\neq c\neq 0,\frac{1}{2},1$ and $2$. Then, we get two unitary matrices
\begin{equation}\label{P1}
P_1=
\left(\begin{smallmatrix}
     0 & 0 & 0 & \frac{1}{\sqrt{2}} & 0 & 0 & 0 & -\frac{1}{\sqrt{2}}\\
     0 & 0 & 0 & 0 & 1 & 0 & 0 & 0\\
     0 & 0 & 0 & 0 & 0 & 1 & 0 & 0\\
     0 & 0 & 0 & 0 & 0 & 0 & 1 & 0\\
     1 & 0 & 0 & 0 & 0 & 0 & 0 & 0\\
     0 & 1 & 0 & 0 & 0 & 0 & 0 & 0\\
     0 & 0 & 1 & 0 & 0 & 0 & 0 & 0\\
     0 & 0 & 0 & \frac{1}{\sqrt{2}} & 0 & 0 & 0 & \frac{1}{\sqrt{2}}
\end{smallmatrix}\right),
\end{equation}
and
\begin{equation}\label{P2}
P_2=\left(\begin{smallmatrix}
     0 & 0 & 0 & 0 & 0 & 0 & -i & 0\\
     0 & 0 & 0 & 0 & 0 & i & 0 & 0\\
     0 & 0 & 0 & 0 & -i & 0 & 0 & 0\\
     0 & 0 & 0 & \frac{1}{\sqrt{2}}i & 0 & 0 & 0 & -\frac{1}{\sqrt{2}}i\\
     0 & 0 & 0 & -\frac{1}{\sqrt{2}}i & 0& 0 & 0 & -\frac{1}{\sqrt{2}}i\\
     0 & 0 & i & 0 & 0 & 0 & 0 & 0\\
     0 & -i & 0 & 0 & 0 & 0 & 0 & 0\\
      i& 0 & 0 & 0 & 0 & 0 & 0 & 0
\end{smallmatrix}\right),
\end{equation}
such that $\rho=P_1\Lambda P_1^{\dag}$ and $\rho'=P_2\Lambda P_2^{\dag}$. Let $P=P_1P_2^{\dag}$, it is easy to verify that $P$ is an unitary matrix and $rank(R(P)_{i|\widehat{i}})=1$ for all $i$. In fact $P=\left(\begin{smallmatrix}
     1 & 0\\
     0 & 1\\
\end{smallmatrix}\right)\otimes\left(\begin{smallmatrix}
     0 & i\\
     i & 0\\
\end{smallmatrix}\right)\otimes\left(\begin{smallmatrix}
     0 & -1\\
     1 & 0\\
\end{smallmatrix}\right)$. Thus $\rho$ and $\rho'$ are LU equivalent.

\section{Conclusions}

We have studied the SLOCC and LU equivalence for generic multi-qubit states 
using a new tensor approach. 
Specifically we associate the coefficient tensors to the quantum states for studying their SLOCC and LU equivalence properties.
Utilizing the mode-n matricization of the coefficient tensors, we have formulated criteria of
SLOCC and LU equivalence for multipartite quantum states. We have adopted the CP decomposition to give
a practical method to uncover the relationship between the tensor and its $n$-mode unfolding and then use the information to characterize the SLOCC and LU equivalence. Detailed examples give a visual display how to identify the equivalence by the new method. However, the $n$-mode matricization of the coefficient tensors for mixed states only gives some necessary conditions under the SLOCC equivalence, namely we can only judge the unequal case. For LU equivalence of mixed states, we derive the necessary and sufficient conditions based on the spectral decomposition and Lemma \ref{3}. The new approach not only provides a new way to detect SLOCC and LU equivalence, but also reveals some characteristic properties of the general multipartite states.

\bigskip

\section*{Data Declaration}

All data generated during the study are included in the article.

\bigskip

\addcontentsline{toc}{chapter}{Acknowledgments}
\section*{Acknowledgment}
The research is partially supported by Simons Foundation grant no. 523868 and National Natural Science Foundation of China under grant nos. 12126351, 12126314 and 11861031. This project is also supported by the specific research fund of the Innovation Platform for Academicians of Hainan Province under Grant No.YSPTZX202215 and Hainan Provincial Natural Science Foundation of China under Grant No.121RC539.

\end{document}